\documentclass[aps,prb,twocolumn,showpacs]{revtex4-1}
\usepackage{graphics}
\graphicspath{{figs/}}
\usepackage{subfigure}
\usepackage{epsfig}
\usepackage{epsf,epic}
\usepackage{color}
\usepackage{marvosym }
\usepackage{subfigure}
\usepackage{siunitx}
\usepackage{amsmath}
\usepackage{amssymb}
\usepackage{amsfonts}
\usepackage{wrapfig}
\usepackage{pstricks}
\usepackage{multirow}
\usepackage{pst-node}
\usepackage{extdash}
\usepackage{bm}
\usepackage{dcolumn}% Align table columns on decimal point
\newcommand{\etal}{\textit{et al.\ }}

\newcommand{\mc}{\multicolumn}
\begin{document}
\title{Quasiparticle self-consistent $GW$ electronic band structures
  of  Be-IV-N$_2$ compounds}
\author{Sai Lyu and Walter R. L. Lambrecht}
\affiliation{Department of Physics, Case Western Reserve University, 10900 Euclid Avenue, Cleveland, Ohio 44106-7079, USA}
\begin{abstract}
  The electronic band structures of BeSiN$_2$ and BeGeN$_2$ compounds
  are calculated using the quasiparticle self-consistent $GW$ method.
  The lattice parameters are calculated for the wurtzite based
  crystal structure commonly found in other II-IV-N$_2$ compounds with
  the $Pbn2_1$ space group. They are determined both in the
  local density approximation (LDA) and
  generalized gradient  approximation (GGA),
  which
  provide lower and upper limits. At the GGA lattice constants, which gives lattice constants  closer to the experimental ones, BeSiN$_2$ is found to have an indirect band gap of 6.88 eV and its direct gap at $\Gamma$ is 7.77 eV, while in BeGeN$_2$ the gap is direct at $\Gamma$ and equals 5.03 eV. To explain the indirect gap
  in BeSiN$_2$ comparisons are made with the parent III-N compound
  w-BN band structure.
  The effective mass
  parameters are also evaluated and found to 
  decrease from BeSiN$_2$ to  BeGeN$_2$. 
\end{abstract}

\maketitle
\section{Introduction}
BeSiN$_2$ and BeGeN$_2$ form part of a larger family of II-IV-N$_2$ nitrides,
which can be viewed as derived from the III-N family by replacing two of the group-III atoms in each tetrahedron surrounding the N by a group-II atom and
two by a group-IV atom. By expanding the family of group-III nitrides including these heterovalent ternaries, significant new  opportunities for band
structure engineering and materials property design are opened. The occurrence
of two different valence cations in these structures presents new challenges
in terms of stoichiometry control, understanding the possible disorder effects,
and more complex defect physics but also offers new possibilities in combination
with existing nitrides exploiting band-offsets between lattice matched
pairs of compounds. 
For an overview of this materials family  and related recent work, see Refs. \onlinecite{Lambrechtbook,Martinez17,ictmc21}.

The  special interest in
Be as group II element arises from it being a second
row element in the periodic table. This makes the related compounds
closely related to BN, the III-N parent compound. BN itself is an interesting
material usually found in a layered form similar to graphite, and knows
as hexagonal or h-BN. However, at high pressures a tetrahedrally bonded
form exists, which is usually found in the cubic zincblende polytype.
Nonetheless there also exists a hexagonal tetrahedrally bonded form
with the wurtzite structure.  Both these forms are considered
extremely hard materials comparable to diamond in their properties.
The atmospheric pressure form however is graphite like and
involves trigonal bonding because of $\pi$-bonding between the
second row elements B and N.  By replacing B with Be and Si or
Be and Ge we may preserve some of the unique properties of
BN and at the same time favor the tetrahedral bonding resulting in
extremely hard and wide band gap materials. 

BeSiN$_2$ and BeGeN$_2$ are expected
to have among the highest band gaps in this family of materials and could
thus be useful for optical devices with wavelengths in the ultraviolet region. 
Synthesis of BeSiN$_2$ and its crystal structure were reported
by Eckerlin \etal in 1967.\cite{Eckerlin67-1,Eckerlin67-2}
The structural properties and electronic band structures at the level of density functional theory (DFT) of BeSiN$_2$ and BeGeN$_2$ were reported by Huang \etal\cite{Huang2001} and Shaposhnikov \etal\cite{Shaposhnikov2008}.
However, DFT in the local density approximation (LDA)
is well known to underestimate the band gaps for semiconductors. A more accurate method is needed to predict  the electronic properties of the Be-IV-N$_2$ compounds, which have not yet been determined experimentally. 
Here we present a study of the band structures using the quasiparticle self-consistent $GW$ method, \cite{MvSQSGW} which provides excellent
agreement with experimental band gaps for most tetrahedrally bonded semiconductors. The goals of this work are to predict accurate band gaps, valence
band fine structure and effective
masses of these materials and to provide and understanding of the direct
or indirect nature of the gaps. 
The details of the computational methodology used here can be found
in section \ref{method}. The structural properties are given in section \ref{str}, and the discussion on stability is in \ref{stability}. The electronic band structures are shown in section \ref{band},  and the effective masses in section \ref{mass}.  We summarize the main results in section \ref{conclusion}.

\section{Computational method}
\label{method}
The structural properties are obtained via the Broyden-Fletcher-Goldfarb-Shanno (BFGS) minimization method \cite{BFGS} using the ABINIT package\cite{abinit} in both local density approximation(LDA) and  generalized-gradient approximation in the Perdew-Burke-Ernzerhof form(PBE-GGA)\cite{PBE} within the general
context of 
density functional theory. The BFGS method allows one to optimize the cell shape and atomic positions simultaneously because
it makes use of the stress tensor and hence gradients are available for the
total energy with respect to atomic displacements as well as the
lattice constants defining the shape and volume of the unit cell. 
The interaction between valence electrons and ions are described
using a pseudopotential.  To be more specific, we used the Hartwigsen-Goedecker-Hutter (HGH) \cite{HGH} pseudopotentials in LDA and Fritz-Haber-Institute (FHI)\cite{FHI} pseudopotentials in GGA. The wave functions are expanded in a plane wave (PW) basis set with energy cutoff of 100 Hartree and a $4\times4\times4$ \textbf{k}-point mesh sampling of the Brillouin  zone is used. The forces are relaxed to be less than $5\times10^{-5}$ Hartree/Bohr.

Next, the density functional theory (DFT) \cite{Kohn-Sham} band structures are calculated using the full-potential linearized muffin-tin orbital (FP-LMTO) all-electron method.\cite{Methfessel,Kotani10,lmsuite} This provides a check
on the pseudopotential band structures with a method
that does not require pseudopotentials. In the FP-LMTO calculations, we used a double-$\kappa$ basis set where $\kappa$ denotes the decay length of the
spherical wave basis function envelopes, which are smoothed Hankel functions.
These are then augmented inside the spheres in spherical harmonics up to
$l=4$ times radial functions and their energy derivatives corresponding
to the actual all-electron potential inside the spheres. 
Specifically, $spd-sp$ on Be atom and $spdf-spd$ on other atoms are included
in the basis set. The  Ge-$3d$ were treated as bands using local orbitals.
The Brillouin zone was sampled using a $6\times8\times8$  \textbf{k}-point mesh.

This DFT band structure is used as the starting point for the 
quasiparticle self-consistent (QS) $GW$ method.\cite{MvSQSGW,Kotani07}
The $GW$ method is a many-body perturbation theoretical method first introduced by Hedin.\cite{Hedin65,Hedin69} In this method, the complex
and energy dependent self-energy $\Sigma(\omega)$
describes the dynamic interaction effects beyond the DFT level
self-consistent field. In the $GW$ approximation it is
given in terms of the one-electron Green's function $G(12)$ and the screened Coulomb interaction, $W(12)$, as $\Sigma(12)=iG(12)W(1^+2)$, where
$1$ is a short hand for the position, spin and time of particle 1 and $1^+$
indicates a time infinitesimally after $t_1$.
The screened Coulomb interaction is itself obtained from the
Green's function via $P(12)=-iG(12)G(21)$ and $W=[1-vP]^{-1}v$, where
$v$ is the bare Coulomb interaction and $W$ the screened one. 
All of these equations are in fact solved after Fourier transformation to {\bf k}-space and energy $\omega$ and in a basis set of auxiliary
Bloch functions, which consists of product functions of muffin-tin orbitals
inside the spheres and interstitial plane waves. But the above real-space
and time notation just provides the most concise way of stating the method's
approximations. Details of the basis set implementation in terms of
muffin-tin orbitals are provide in Ref.\onlinecite{Kotani07}

In the QS$GW$ method,\cite{Kotani07}
the energy dependent self-energy matrix in the
basis of the independent electron Hamiltonian (LDA)
eigenstates is replaced by an energy averaged and Hermitian matrix
\begin{equation}
\tilde \Sigma_{ij}=\frac{1}{2}\mathrm{Re}\{ \Sigma_{ij}(\epsilon_i)+\Sigma_{ij}(\epsilon_j)\}
\end{equation}
This then replaces the starting LDA (or GGA) exchange-correlation potential
and defines a new independent particle Hamiltonian $H_0$ whose eigenvalues
and eigenstates provide a new $G_0$
and through the next $G_0W_0$ calculation the next $\tilde\Sigma$.
The process is then iterated till the
self-energy $\tilde \Sigma$ is self-consistent at which point the Kohn-Sham
eigenvalues of $H_0$ coincide with the quasiparticle excitation energies
of the many-body theory. Hence the name quasiparticle self-consistent $GW$.
Note that in the process the eigenfunctions are recalculated in
each step by rediagonalizing the $H_0$ of that iteration which
includes the $\tilde\Sigma$ instead of the original $v_{xc}$ from the
previous step.  The details of the implementation of the
QS$GW$ method  and various technical aspects
can be found in Ref. \onlinecite{Kotani07}. 
In the present QS$GW$ calculation, the $\Sigma(\omega) $ is calculated
up to $\omega=2.5$ Ry, and is approximated by a diagonal average matrix when it is above 2 Ry before calculating the quasiparticle shifts in
the next step.  A cut-off of 3.5 Ry is used for the interstitial plane-waves
and for the Coulomb interactions auxiliary basis set.  The \textbf{k}-point mesh over which the $GW$ self-energy is evaluated was set as $3\times3\times3$.
The $\Sigma$ is at this point known as a matrix in the basis of $H_0$
eigenstates $\psi_{n{\bf k}}$.  Expanding these in the basis set of 
muffin-tin orbital Bloch functions
$\chi_{RL{\bf k}}$, and subsequently each of these into
  $\chi_{RL{\bf T}}$ by an inverse Bloch sum (or discrete Fourier transform)
the $\tilde \Sigma_{RL,R'+{\bf T}L'}$ is then obtained in real space up to a
certain range. Here, $R$ denotes the site in the unit cell, $L$ the angular
momentum and other indices needed to specify the specific basis function
centered at this site, and ${\bf T}$ the
lattice vectors. This can now be  converted
back by a direct Bloch sum to the $6\times8\times8$ \textbf{k}-point set used for the charge self-consistency iterations of $H_0$ and eventually also to the \textbf{\bf k}-points
along symmetry lines of the Brillouin zone for the band structure plots.
This effective interpolation scheme allows us to obtain the bands at the $GW$ level at any \textbf{k}-point and hence also accurate effective masses.
It is essentially equivalent to a Wannier function interpolation scheme,
except that in the LMTO approach we do not need to construct
Wannier functions because we already have a relatively 
short-range  atom centered basis set. 

Finally, we note that in the QS$GW$ approximation, the screening of the
Coulomb interactions is usually found to be underestimated by
about 20 \%,\cite{Deguchi16,Churna18}
resulting in band gaps being overestimated. This can
be corrected by mixing 80\% of the $\tilde\Sigma$ with 20 \% of the original
LDA $v_{xc}^{LDA}$ and is referred to as the $0.8\Sigma$ approximation.
We note that this may sound similar to a hybrid functional but is in practice
quite different because there is no adjustment of this fraction involved
and this represents in effect going beyond standard $GW$ by including
additional diagrams in the calculation of $W$ which could now in fact
be done explicitly following  the Bethe-Salpeter Equation (BSE)
approach,\cite{Cunningham18} but at a much higher cost than the standard
QS$GW$ + 80 \%
renormalization. In other words, where the calculations have been done
at this beyond standard $GW$ level, for example using a time dependent
DFT (TDDFT) approach with an exchange correlation kernel in the screening
of $W$ extracted from BSE,\cite{Shishkin07,PasquarelloChen15}
they tend to  confirm the 80\% renormalization approach as analyzed
in Ref. \onlinecite{Churna18}. 
\section{Results}

\subsection{Crystal structure}
\label{str}

\begin{table}
\centering
\caption{Calculated lattice parameters of Be-IV-N$_2$ compared with experimental values.\label{tablattice}}
\begin{ruledtabular}
\begin{tabular}{lccccccc}
Compound &method& $a $(\AA) & $ b$(\AA)  &  $c $(\AA)  & $V$(\AA$^3$) & $b/a_{w}$ & $c/a_{w}$ \\ \hline
 BeSiN$_{2}$ & LDA&5.67 & 4.92 & 4.62 & 128.9 & 1.74 & 1.63 \\
                          & GGA&5.75 & 4.98 & 4.68 & 134.0 & 1.73& 1.63 \\
                          & Expt.\footnote{From Eckerlin\cite{Eckerlin67-2}}&5.747 & 4.977 & 4.674 & 133.7 & 1.73 & 1.63\\
                          &LDA\footnote{From Shaposhnikov \etal \cite{Shaposhnikov2008}} & 5.697 & 4.939 & 4.639 &  133.7 & 1.73 & 1.63 \\
                          &GGA $^b$& 5.772 & 4.999 & 4.699 & 135.6  & 1.73 & 1.63 \\
\\
 BeGeN$_{2}$ & LDA&5.81 &5.07 & 4.77 & 140.5 & 1.75 & 1.64 \\
                            & GGA&5.93 & 5.17 & 4.86 & 149.0 & 1.74 & 1.64 \\
                            &LDA$^b$ & 5.856 & 5.105 &4.803   &143.6 & 1.74 & 1.64  \\
                            &GGA$^b$ & 5.972 & 5.204 & 4.894  & 152.1 & 1.74 & 1.64    \\

\end{tabular}
\end{ruledtabular}
\end{table}

\begin{figure}
\includegraphics[width=11cm]{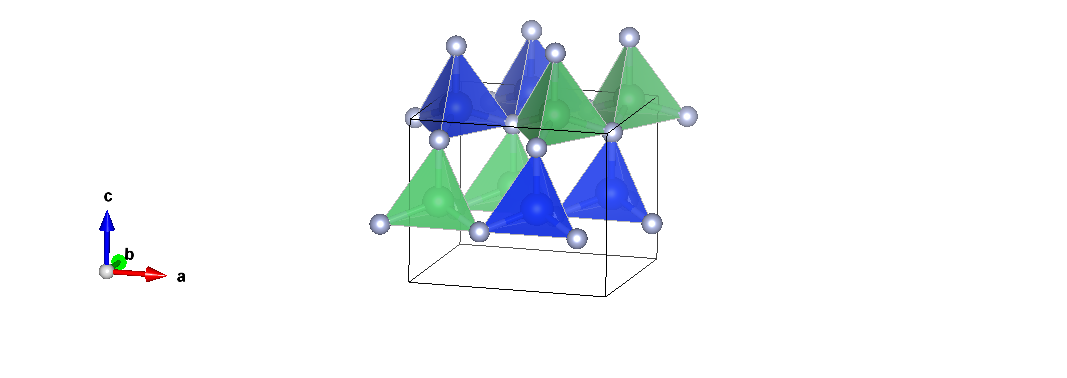} \vskip -0.0 cm
\caption{Crystal structure of BeSiN$_2$.The small open spheres represent
  N atoms, the green and blue larger spheres inside the nearest neighbor
  tetrahedrons represent Be and Si respectively.\label{besin2cryst}}
\end{figure}

The calculated lattice constants for BeSiN$_2$ and BeGeN$_2$ in both LDA and GGA are given in Table  \ref{tablattice}. The space group is $Pbn2_1$ and
the order of the lattice constants is chosen so that $a>b$.  The relation of the lattice constants in the idealized orthorhombic structure to those in
the wurtzite structure is given by $a=2a_w$, $b=\sqrt3a_w$ and $c=c_w$.
The optimized $b/a_w$ ratio indicates how far the structure is from
the idealized wurtzite structure. From the results
in Table \ref{tablattice} we can see that BeSiN$_2$ is closer to the idealized
wurtzite structure than BeGeN$_2$. The GGA lattice constants are systematically larger than the LDA ones which is commonly found.  We also compare our relaxed lattice constants with those reported in literature and with the experimental
values for BeSiN$_2$. Good agreement is found.
Only the experimental lattice constants of BeSiN$_2$ are available. In this case, we can see that GGA lattice constants are more accurate than LDA ones. Both LDA and GGA produce nearly the same $b/a_w$ and $c/a_w$ ratios in both
materials studied here. The crystal structure of BeSiN$_2$ is shown in Fig. \ref{besin2cryst} and BeGeN$_2$ has a similar structure.

The atomic positions of the atoms in the unit cell are given in terms of Wyckoff 4a positions in Table \ref{tabwyckoff}. The origin is set at the position of the 2-fold screw axis in the $xy=ab$ plane. We may note that the nitrogen atoms are sitting nearly on top of the Be and group-IV atoms which  shows that BeSiN$_2$ has its orthorhombic crystal structure very close to the supercell of wurtzite
without much distortion. 

\begin{table}
\centering
\caption{Wyckoff positions of Be-IV-N$_2$ in the GGA approximation .\label{tabwyckoff}}
\begin{ruledtabular}
\begin{tabular}{lcccc}
Compound &atom& $x $ & $ y$ &  $z $  \\ \hline
 BeSiN$_{2}$   
   &Be&0.125 & 0.083 & -0.013\\
    &Si&0.625 & 0.086 & -0.014\\
   &N$_\mathrm{Be}$&0.127 & 0.084 & 0.365\\
  &N$_\mathrm{Si}$ & 0.623& 0.080 & 0.362\\
\\

 BeGeN$_{2}$ 
   &Be&0.125 & 0.080 & -0.013\\
    &Ge&0.624 & 0.091 & -0.013\\
    &N$_\mathrm{Be}$&0.115& 0.072 & 0.351\\
  &N$_\mathrm{Ge}$ & 0.634& 0.091 & 0.375\\
\end{tabular}
\end{ruledtabular}
\end{table}

\subsection{Formation energy and stability}
\label{stability}
We next check the  thermodynamic stability. The energies of formation
are calculated from the cohesive energies as given in Table \ref{formation}
which are calculated in the GGA approximation by subtracting the atomic
energies from the solid's total energy. The atomic energies include the spin-polarization energy. The N$_2$ molecule was calculated in a large unit cell
using the FP-LMTO method and including additional augmented plane wave basis functions\cite{Kotani10}  to well represent the region outside the molecule, which is important
to obtain total energies which are not sensitive to the choice of
muffin-tin radii. The energies of formation
of both compounds are negative, meaning that they are stable relative to
the elements in their respective phases at standard conditions (room
temperature and atmospheric pressure).  They are also found to be stable
relative to the competing binary compounds. The formation energies of
the latter were calculated for the phases already found to be the minimum
energy structures in the Materials Project \cite{mp} and compared
with the data in that database in Table \ref{formation}.
They show that the BeSiN$_2$ is stable against the reaction
\begin{equation}
  3\mathrm{BeSiN}_2\rightarrow \mathrm{Be}_3\mathrm{N}_2+\mathrm{Si}_3\mathrm{Si}_4
\end{equation}
by 0.136 eV/atom and similarly for the Ge case by 0.176 eV/atom.

\begin{table}
\centering
\caption{Cohesive energies and formation energies (in eV/atom)
  calculated in the PBE-GGA approximation.\label{formation}}
\begin{ruledtabular}
\begin{tabular}{lcc}
    &  ours  & MP\footnote{Material Project\cite{mp}} \\ \hline
Be  & 3.44                    \\
Si  & 4.49\\
Ge & 3.39 \\
N  & 5.17\\        \\
Be$_3$N$_2$  & -1.24 & -1.225  \\
Si$_3$N$_4$    & -1.04  & -1.312 \\
Ge$_3$N$_4$   & -0.08  & -0.26 \\
BeSiN$_2$   &-1.26   &  -1.412   \\
BeGeN$_2$ & -0.74  & -0.813   \\

\end{tabular}
\end{ruledtabular}
\end{table}

\subsection{Energy bands and density of states}
\label{band}

\begin{figure*}
\includegraphics[width=12cm]{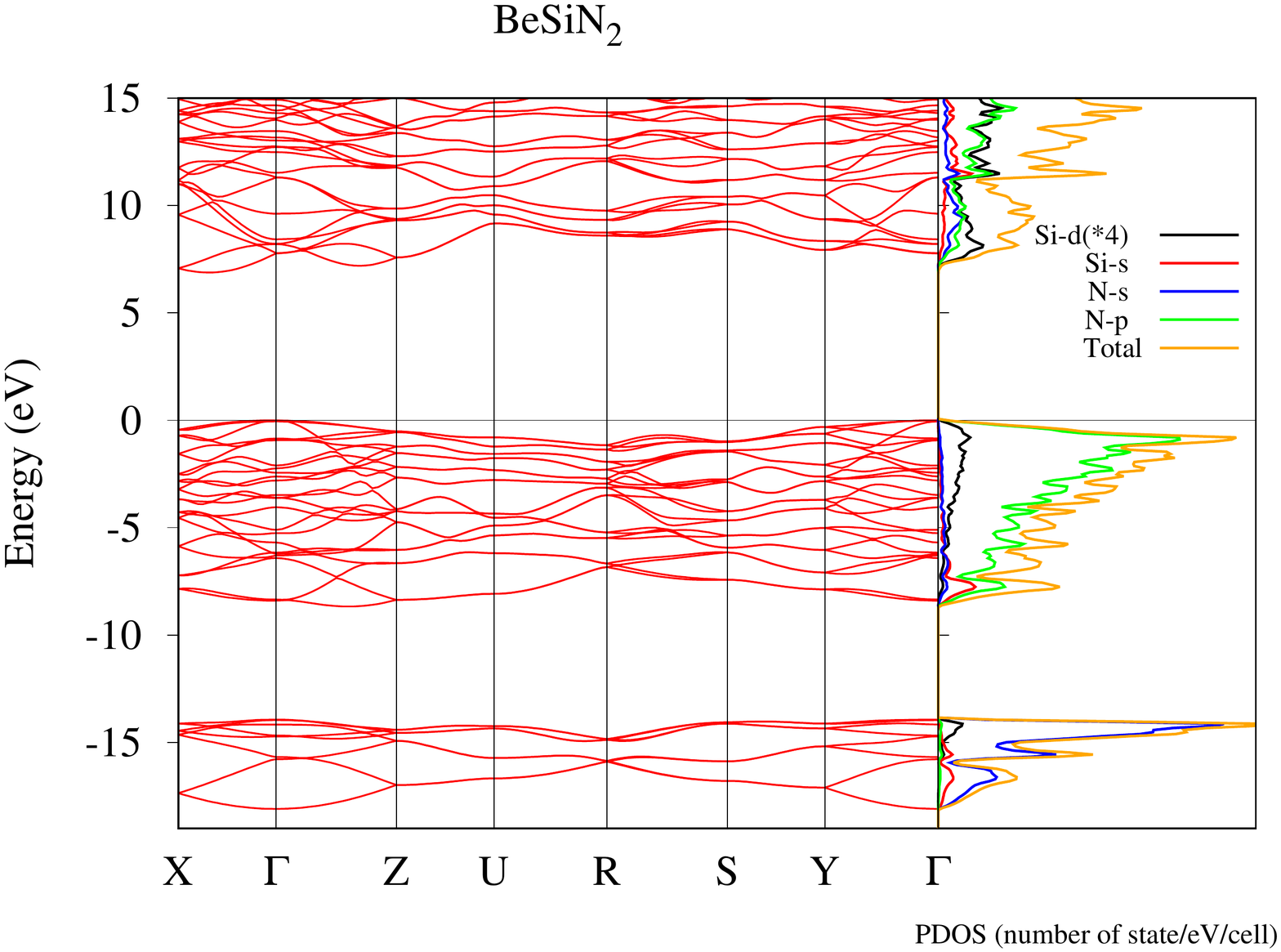} 
\includegraphics[width=12cm]{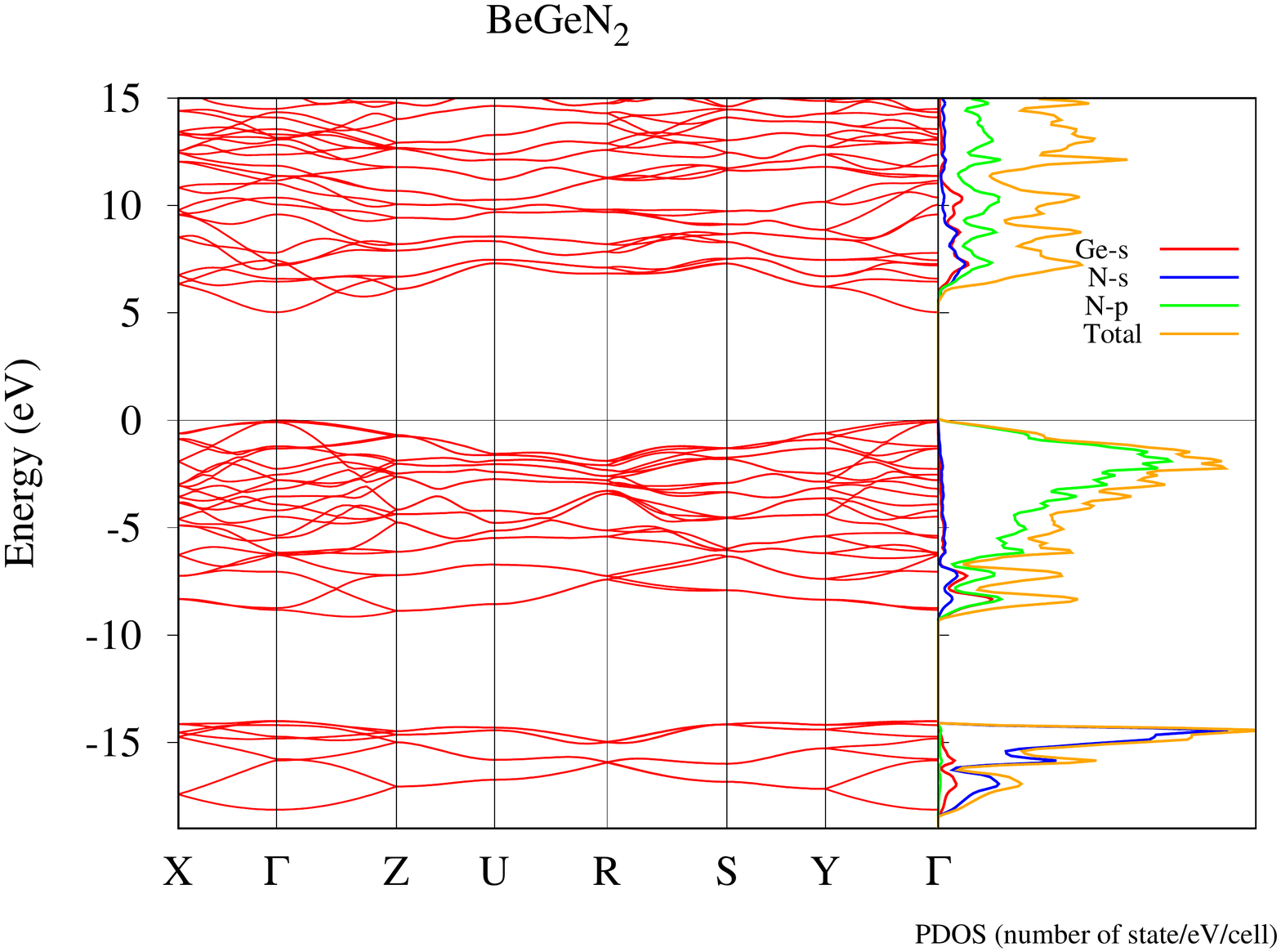} 
\caption{Band structures and partial densities of states
  of BeSiN$_2$ and BeGeN$_2$ in the QS$GW$ approximation at the GGA lattice constants.\label{gwband}}
\end{figure*}

\begin{figure}
\includegraphics[width=8.5cm]{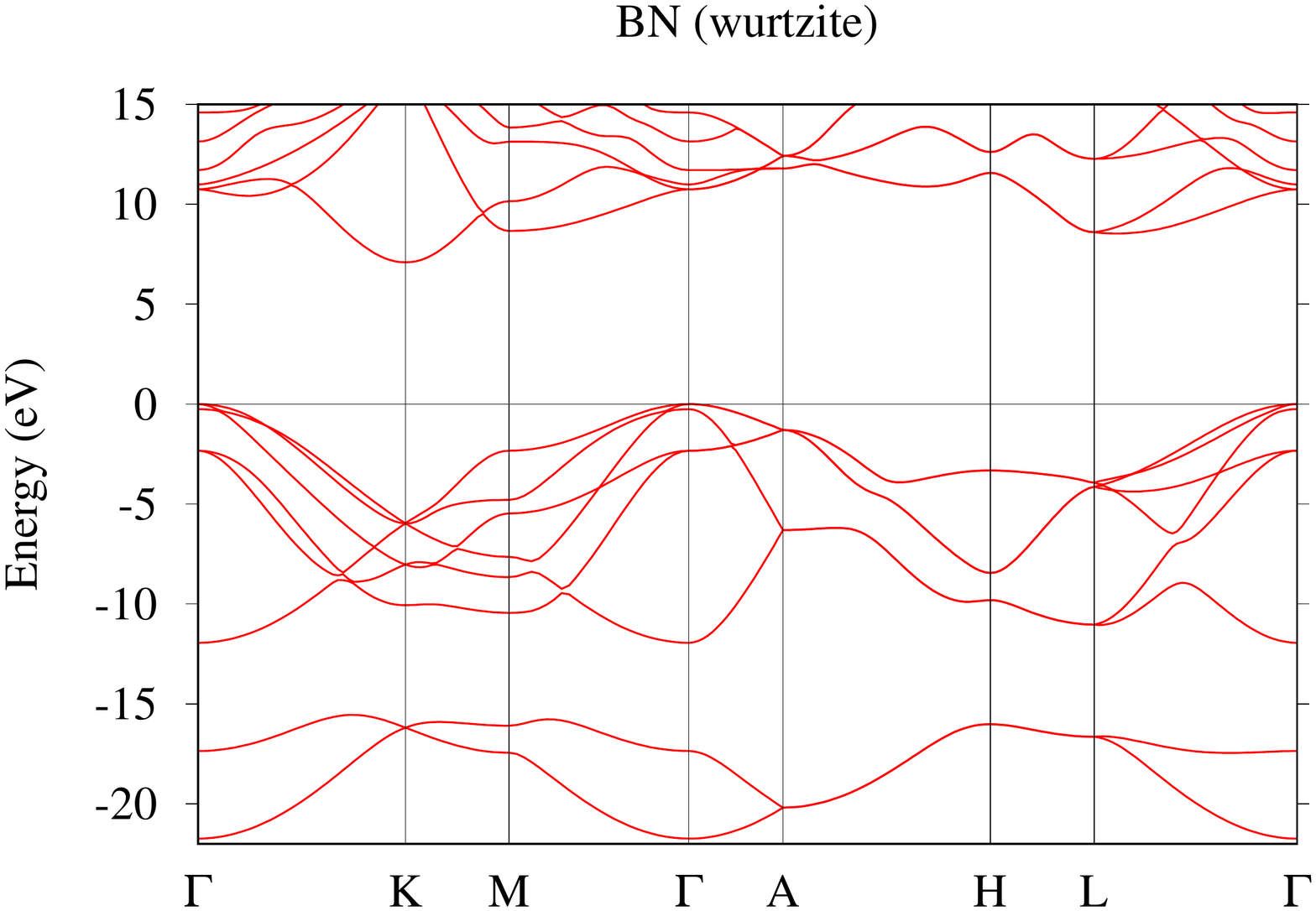}  \vskip -1cm
\includegraphics[width=8.5cm]{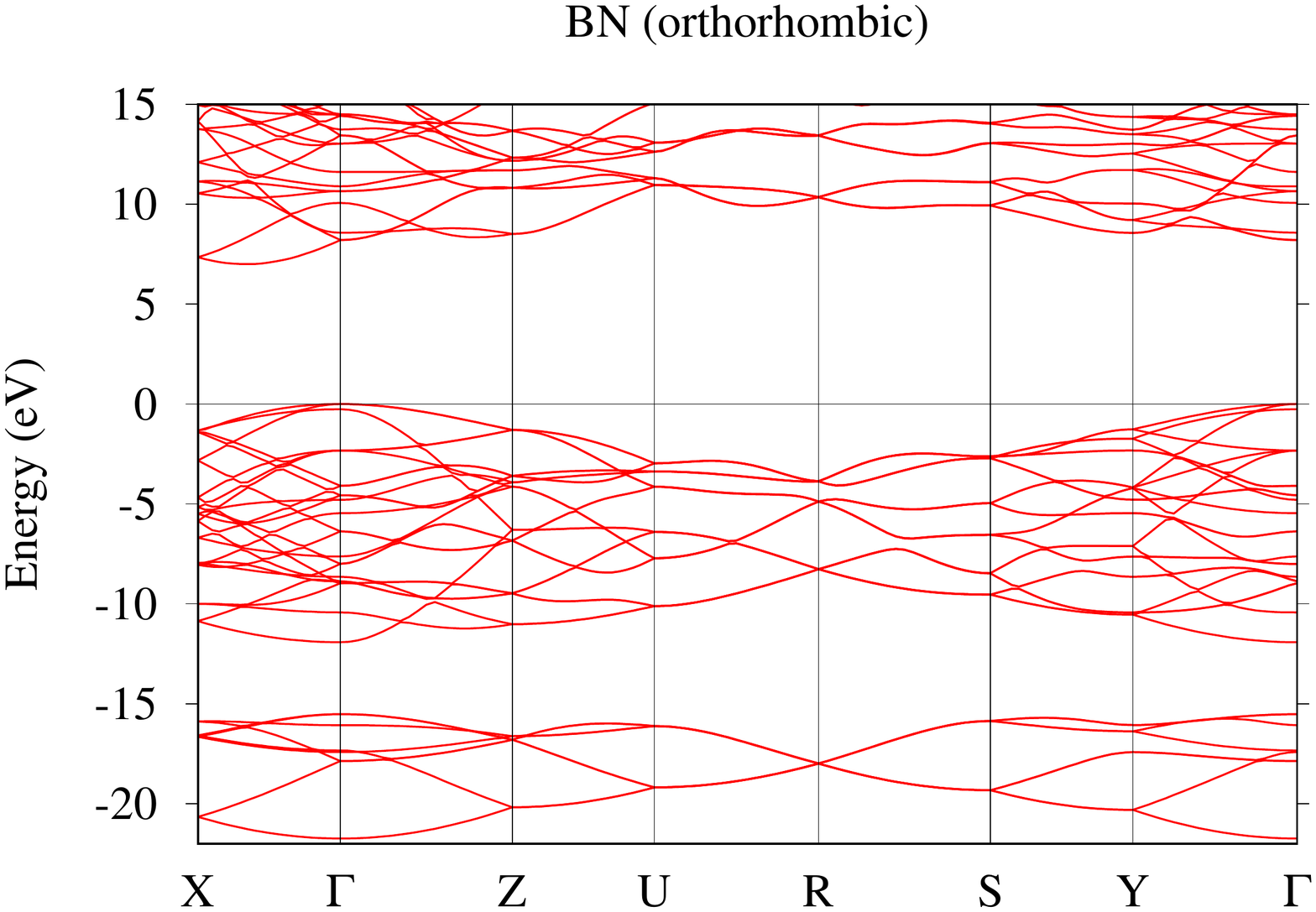}  \vskip -1cm
\caption{Band structures of BN plotted in Brillouin zone of wurtzite and orthorhombic structure in the QS$GW$ approximation. \label{bnorth}}
\end{figure}

\begin{figure}
\vskip -1 cm\includegraphics[width=8cm]{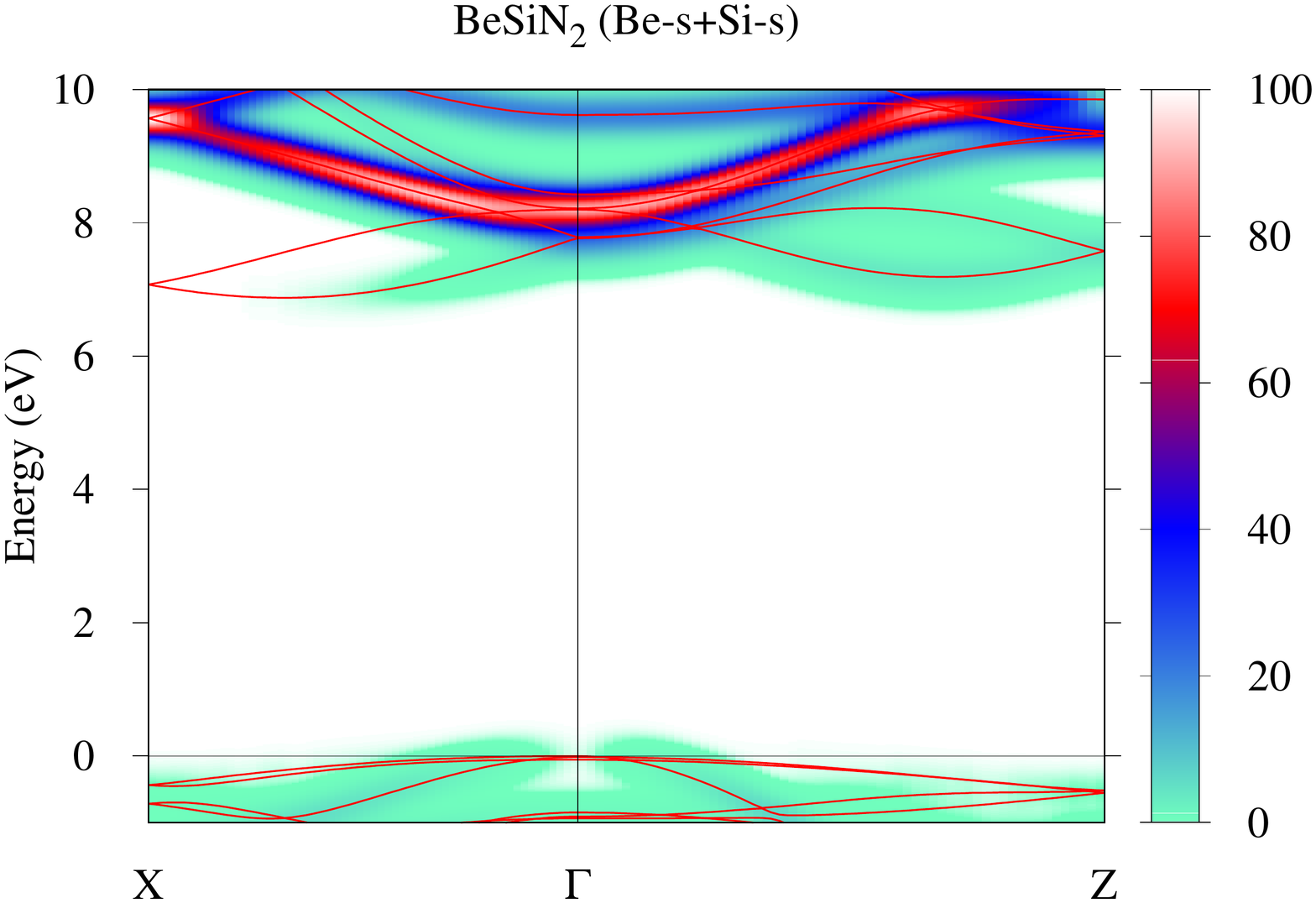}
\vskip -1 cm\includegraphics[width=8cm]{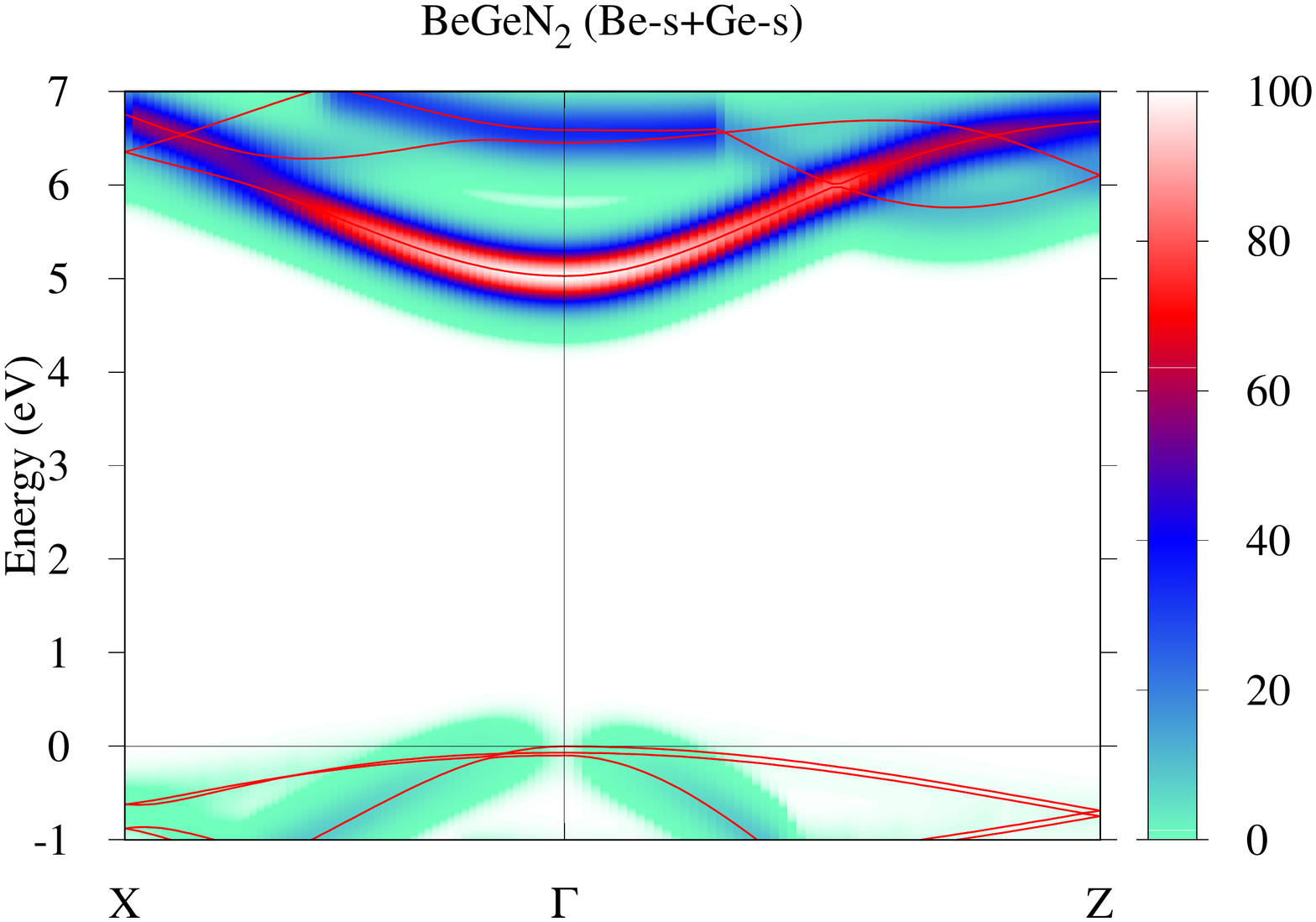} 
\caption{QS$GW$ band structures near CBM projected on
  the specified muffin-tin orbital basis functions.\label{colorband}}
\end{figure}

The band structures and partial densities of states (PDOS)
in the 0.8$\Sigma$ QS$GW$ approximation at GGA  lattice constants are shown in Fig. \ref{gwband}. The PDOS show that, as expected, the lowest
set of 8 valence bands near $-15$ eV are dominated by N-$s$ and the
next set of valence bands (between $-9$ eV and the VBM set at 0),
by N-$p$ orbitals. The conduction bands
have more cation character  but also N-$s$ and N-$p$
because the cation states are antibonding with the N-orbitals.
In the case of BeSiN$_2$,
we see it has even a strong Si-$d$ character. The conduction band
orbital character will be analyzed in more detail
in a {\bf k}-resolved manner below.  

BeGeN$_2$ is found to be a direct band gap semiconductor while BeSiN$_2$ has an indirect band gap. In BeGeN$_2$, both the CBM and VBM occur at the $\Gamma$ point. An indirect band gap was previously also found for ZnSiN$_2$,\cite{punya11} MgSiN$_2$\cite{Atchara16} and CdSiN$_2$\cite{Sai17}. However, in those cases, the CBM is at $\Gamma$ and the VBM near one
of the Brillouin zone edges. On the other hand, in BeSiN$_2$,  the
VBM is at $\Gamma$ but the CBM is located between the $\Gamma$ and $X$ points.
In Ref. \onlinecite{Shaposhnikov2008}, the CBM for BeSiN$_2$ occurs between $\Gamma$ and $Y$ which is  simply because of the difference in labeling. Their $Y$ is actually labeled as $X$ in the present work because the $a$ and $b$ axes are
interchanged. This location of the CBM is robust and in agreement
between LDA, GGA and QS$GW$ calculations. The CBM position is at 2/3 of
the $\Gamma-X$ line irrespective of whether
we use the LDA or GGA lattice constants.

To further investigate the origin of the indirect gap in BeSiN$_2$, 
we compare it with the band structure of wurtzite BN, which is shown in
Fig. \ref{bnorth}.
Note that wurtzite BN is the most closely related
III-N parent compound of the II-IV-N$_2$
compounds considered here because B is adjacent to Be in the periodic table. 
We first show it in the standard wurtzite hexagonal Brillouin zone and
then in the Brillouin zone  corresponding to the
orthorhombic $2\times\sqrt{3}\times1$ supercell for easier comparison
to the II-IV-N$_2$ materials whose primitive cell corresponds to
this supercell of the underlying wurtzite type lattice. 

We can understand these relations in terms of band folding.
The relation between the hexagonal Brillouin zone and
the orthorhombic Brillouin zone is shown in Fig. 3 of Ref.
\onlinecite{Lambrechtzgn05}.
The wurtzite $\Gamma-M$ line is folded in two along the $y$-direction
and becomes the $\Gamma-Y$ direction in the orthorhombic structure. The
$\Gamma-K$ is also folded along the $x$ direction but the point $K$
lies at  $4\pi/3a_w$ along the $x$ direction and the orthorhombic $X$
about which we fold the bands lies at $\pi/2a_w$.  
So, the wurtzite w-BN conduction band minimum which occurs at $K$ of the
hexagonal Brillouin zone ends up at 2/3 $\Gamma-X$ in the orthorhombic
Brillouin zone. This is very close to the CBM location also in BeSiN$_2$.  

The band structure of w-BN in the CB is rather complex when folded
in the orthorhombic zone and is itself not so well known because
w-BN is a rare form of BN. In its tetrahedrally bonded modification 
which occur at high pressure, the zincblende form  has lower energy
than the wurtzite. So, we next address the relation between zincblende
and wurtzite BN.  First, in z-BN, the CBM occurs at $X$ of the fcc Brillouin
zone and in w-BN it occurs at $K$ of the hexagonal Brillouin zone.
Both of these states are mixtures of $s$ and $p$-like atomic orbitals
and also have a strong component in the interstitial region, in particular
the large open channel of wurtzite. A similar situation has
recently been pointed out for SiC.\cite{Matsushita12} But here we
are concerned with the states at $\Gamma$. 
Even in z-BN the $p$-like $\Gamma_{15c}$ lies below the $s$-like
$\Gamma_{1c}$. In wurtzite,
the $\Gamma_{15}$ splits into $\Gamma_5$ ($x,y$-like) and $\Gamma_1$ ($z$-like)
states. So at $\Gamma$ in w-BN, the conduction band ordering of states is
$\Gamma_5$, $\Gamma_1$, and only the third band is the $s$-like $\Gamma_1$. 
Now after folding into the orthorhombic BZ, the lowest three conduction bands
at $\Gamma$ all arise from the folding of the $\Gamma-K$ and $\Gamma-M$
bands, the next two, a degenerate and nondegenerate band near about 10-11 eV
are derived from the wurtzite $\Gamma_5,\Gamma_1$ B-$p$-like states and
the band a little below 12 eV is the B-$s$-like state.

Comparing this
with BeSiN$_2$, it becomes clear  that here the $\Gamma_5,\Gamma_1$
derived states lie higher than the $s$-like $\Gamma_1$. This is because
the Si-$p$ like states lie at higher energy than B-$p$ like states. 
The CBM at $\Gamma$
is a folded state but the next band already has a strong $s$-like character
on both Be and Si. 
This can be seen in Fig. \ref{colorband}.  In this figure we show the band
colored according to their weight on muffin-tin orbital
basis functions of specific angular momentum character centered on
the different atoms. We recognize the typical strongly dispersing
mixed cation $s$-like conduction band in both materials, but while this is
the lowest conduction band in BeGeN$_2$, it lies just a bit higher
in the BeSiN$_2$ case above the folded bands discussed above. 
Additional colored band structures reflecting their other
atomic orbital characters are given in Supplementary Material.\cite{suppmat}
The CBM has contributions from Be-$p$, Si-$p$ antibonding with N-$s$
and N-$p$. This is also clear from the PDOS except that the
{\bf k}-resolved analysis of the orbital character provides more
detailed information. So, in summary, the reason for the indirect band
structure and the CBM location in the Brillouin zone
is explained by its relation to w-BN parent compound of which
we can view BeSiN$_2$ to be a perturbation. In turn the w-BN
band structure is related to that of z-BN  by band folding effects.
The ordering of these bands depends sensitively on the ordering of
atomic $s$ and $p$-orbitals in the second row of the periodic table
elements Be an B and the larger splitting of $s$ and $p$ valence orbitals in
Ge than Si.

In BeGeN$_2$, the conduction band has a stronger dispersion near its minimum at $\Gamma$, while the valence bands are quite flat. Nonetheless the lowest
conduction band dispersion near its minimum  
is less pronounced than in MgGeN$_2$, ZnGeN$_2$ or CdGeN$_2$.  This is because
the Be-$s$ states lie less deep on an absolute scale.

\begin{figure}
\includegraphics[width=8.5cm]{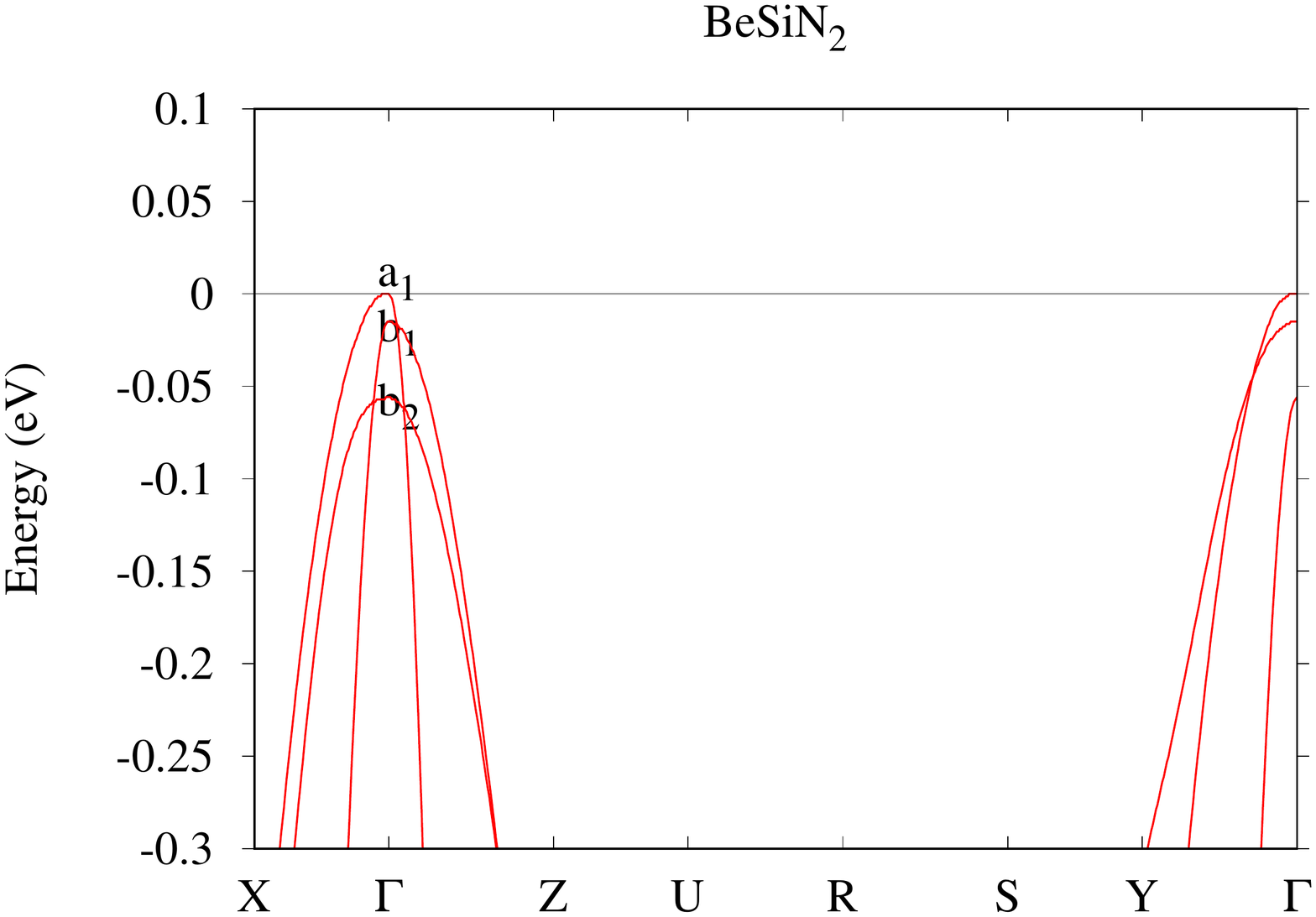}
\includegraphics[width=8.5cm]{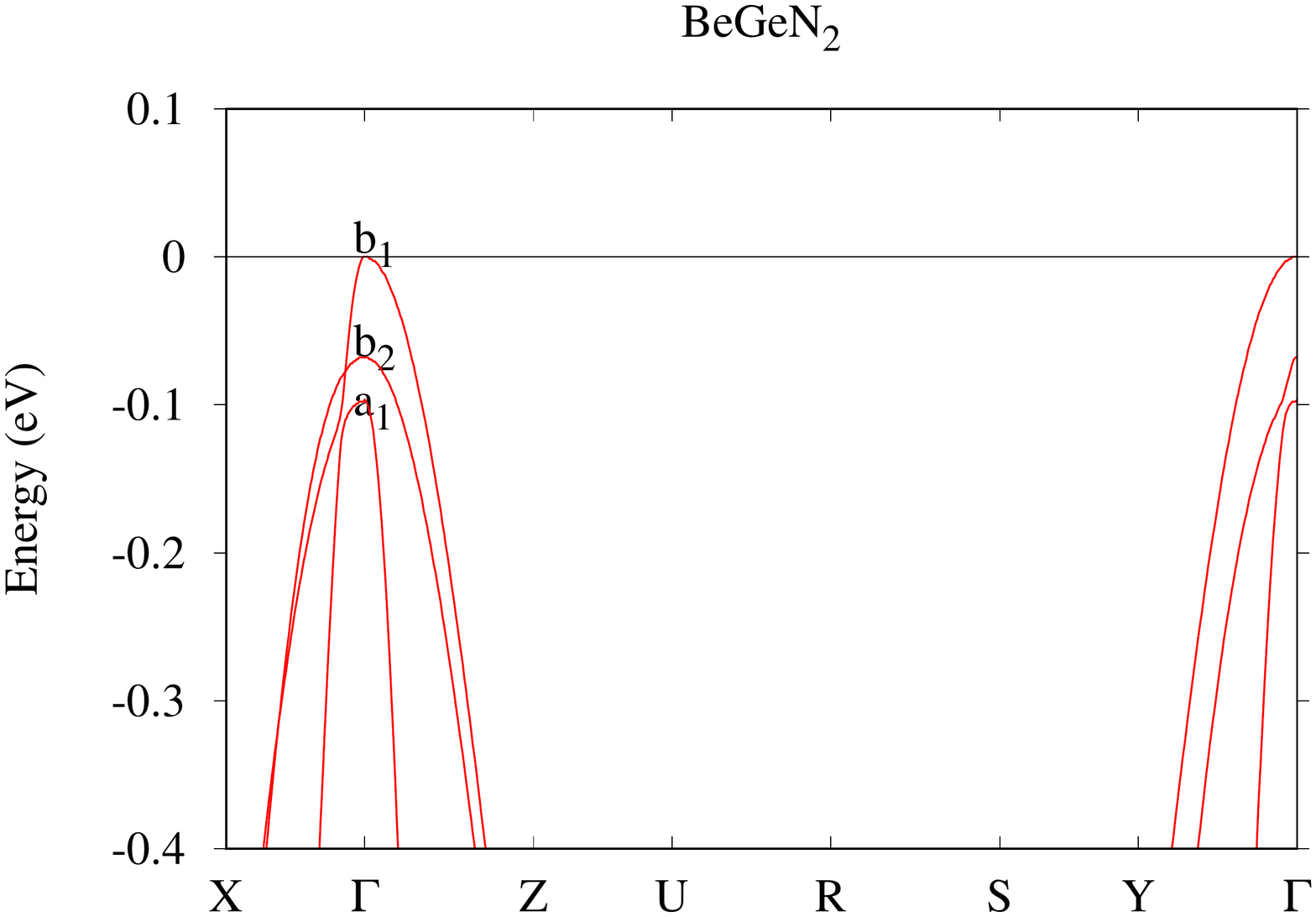} 
\caption{Band structures near VBM region with symmetry labeling at $\Gamma$. \label{bandzoom}}
\end{figure}

A zoom-in  of the electronic band structure near the VBM region for BeSiN$_2$ and BeGeN$_2$ is shown in Fig. \ref{bandzoom}. Here, the energy levels at $\Gamma$ point are labeled by their related irreducible representation of point group $C_{2v}$. The states $a_1$, $a_2$, $b_1$, $b_2$ correspond to basis function $z,xy,x,y$, respectively. These symmetry labels are helpful to understand whether the optical  transitions from the valence bands to the conduction band minimum at $\Gamma$ point are dipole-allowed or not.  In the case of BeGeN$_2$, the CBM has $a_1$ symmetry, so the dipole-allowed optical transitions are from $a_1$ valence states when $E || c$, $b_1$ when $E || a$, and $b_2$ when $E || b$, respectively.  The crystal field splitting of the energy levels near the VBM at
the $\Gamma$ point are given in Table \ref{tabzoomlevel}. In both cases, the top three valence bands have $a_1, b_1$ and $b_2$ symmetries but they
occur in different order in the two compounds.   The dispersion of this manifold of three bands near the VBM
can be represented by a Kohn-Luttinger type Hamiltonian
further discussed in the next section.

The typical automated plotting program only connects the bands
according to their band number in terms of increasing energy
does not take into account which bands are allowed to cross or not.
Therefore to obtain the correct band connectivity a very fine mesh
is needed. We here used about 100 points along each direction. 
We can somewhat further inspect 
which bands cross or have avoided crossings. For example, for ${\bf k}$
along the $\Gamma-X$ direction, the group of ${\bf k}$ only contains
the identity and the $xz$ mirror plane. Now, $b_1$  and $a_1$ are
even with respect to this mirror plane but $b_2$ is odd. Therefore the
band emanating from $b_2$ can cross the one from $b_1$ but the ones
from $b_1$ and $a_1$ have an avoided crossing because
they are allowed to interact having he same even symmetry. Similarly
along $\Gamma-Y$, the remaining group consists of the identity and the
$yz$ mirror plane. With respect to this mirror plane $a_1$ and $b_2$ are
even but $b_1$ is odd. Therefore we see that the $b_2$ and $a_1$ band
do not cross.

The band gaps of BeSiN$_2$ and BeGeN$_2$ calculated by DFT and 0.8$\Sigma$ QS$GW$ at LDA and GGA lattice constants are summarized in Table \ref{tabgap}.
For comparison, the band gaps in w-BN are 7.09 eV (indirect $\Gamma-K$) and
10.76 eV (direct at $\Gamma$).
For BeGeN$_2$, the change in the gap from LDA to GGA lattice constants is as expected for tetrahedrally bonded materials: larger lattice constants relaxed in GGA gives rise to smaller band gaps. The change in gap with respect to the unit cell volume can be quantified by the deformation potential, $a_v=d\ E_g/d\ ln\ V$. The deformation potential of  BeSiN$_2$ is $-2.9$ eV from finite difference calculation, and for BeGeN$_2$ is $-9.6$  eV. The large difference between the two
reflects the fact that the orbital character of the CBM is different in both
materials.  In BeGeN$_2$, we see that the smaller lattice volume of LDA
gives rise to a larger gap already at the DFT level and this is maintained
in the QS$GW$ approximation. 
For BeSiN$_2$, we give both the lowest indirect gap and the direct gap at
$\Gamma$.  In this case, surprisingly, the QS$GW$ gaps at the
GGA lattice constants are slightly larger than those
at the  LDA lattice constants, even though  the lattice constants are larger
in GGA. This results from the different orbital character of the CBM
in this material. Apparently the deformation potentials of the indirect CBM
and even the CBM at $\Gamma$ states
are different from  the usual $\Gamma_1$ (or $a_1$-symmetry) CBM in BeGeN$_2$
or other II-IV-N$_2$ materials. 

\begin{figure}
\includegraphics[width=9cm]{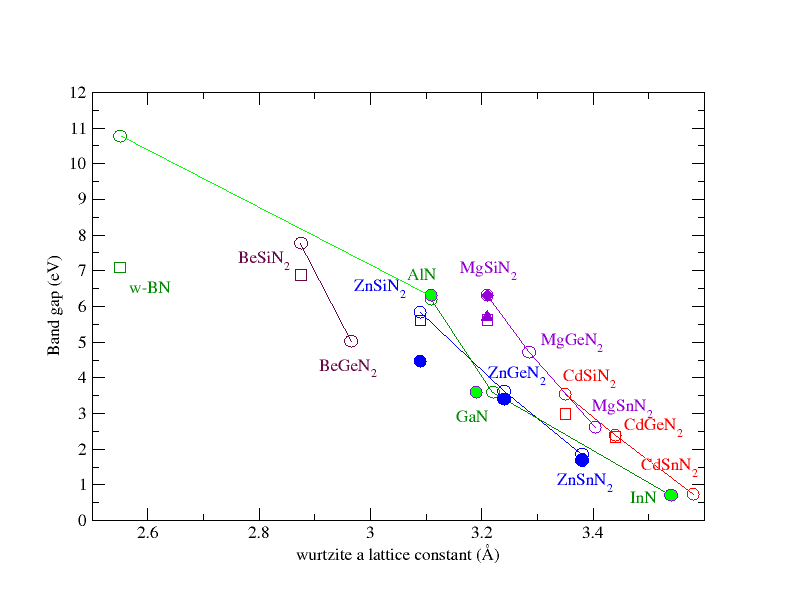} \vskip -0.0 cm
\caption{Band gaps of II-IV-N$_2$ and III-N compounds. The open and filled
  symbols represent our calculations and experiment respectively, except
  for MgSiN$_2$ where the closed symbol refers to a hybrid functional
  calculation.\cite{Quirk14} The circles represent lowest direct gaps
  at $\Gamma$, the squares
  indirect gaps. 
  \label{gaps24n2}}
\end{figure}

To compare the band gaps of these materials with related III-N and other
II-IV-N$_2$ semiconductors, it is useful to show them in a band
gap versus lattice constant diagram as shown in 
in Fig. \ref{gaps24n2}. The lines here are just guides to the eye and do not
include alloy band gap bowing effects.
We can see from this plot that these materials fall somewhere in between
w-BN and  the other III-N or II-IV-N$_2$ materials. 

The direct band gap of BeGeN$_2$ of 5.03 eV corresponds to a wave length of
the emitted or absorbed light of 246 nm in the UV region. The gap is still
somewhat lower than in AlN (6.3 eV) but still possibly useful to
push LEDs toward deep UV compared with GaN or ZnGeN$_2$. The gap
however is only slightly larger than in MgGeN$_2$, another direct gap
material but occurs at a much smaller lattice constant, so it will be
more difficult to integrate with GaN  or ZnGeN$_2$ in heterostructures. 
The direct band gap of BeSiN$_2$ of
7.77 eV corresponds to 159 nm and is even
higher than that in AlN but the indirect nature makes it somewhat
less attractive for such applications.  It might still be a  useful
material for UV detectors.

\begin{table}
  \caption{Energy levels  in meV at $\Gamma$ relative to the VBM at $\Gamma$
    including their symmetry label.\label{tabzoomlevel}}
  \begin{ruledtabular}
    \begin{tabular}{cdcd}
      \mc{2}{c}{BeSiN$_2$}& \mc{2}{c}{BeGeN$_2$} \\
      Sym.&\mc{1}{c}{$E$}&Sym.&\mc{1}{c}{$E$} \\ \hline
        $a_1$     &0   & $b_1$& 0    \\
        $b_1$     &-15.0  & $b_2$& -68.0   \\
        $b_2$     & -55.8  & $a_1$&  -96.6  \\
        $b_2$     & -846.3  & $b_2$& -1206.9   \\
        $a_2$   &  -908.9 & $a_2$& -1311.6   \\
    \end{tabular}
  \end{ruledtabular}
\end{table}

\begin{table}
  \caption{Band gaps (in eV) calculated in the 0.8$\Sigma$ QS$GW$
    and DFT approximations at the LDA and GGA lattice constants.
    The second column indicates at which level of theory the band structure is calculated, the third at which level the lattice constant is minimized. \label{tabgap}}
\begin{ruledtabular}
\begin{tabular}{lcccc}
      &&\mc{2}{c}{BeSiN$_2$  } &BeGeN$_2$  \\
      && indirect & direct & direct   \\ \hline
Ref. \onlinecite{Huang2001}  & DFT LDA & 5.08 & 5.7\footnote{estimated from the band structure figure} & 5.24\footnote{Their lattice constants are smaller than that in this work}   \\ \\
Ref. \onlinecite{Shaposhnikov2008}&DFT  LDA & 4.95 & 5.82 & 3.69 \\
         &DFT GGA & 5.19 & 5.93  & 3.37 \\  \\
This work & DFT LDA & 4.96 & 5.80 & 4.12 \\
                   & DFT GGA & 5.13 & 5.95 & 3.45 \\
                   & GW LDA & 6.82   &7.74 & 5.88  \\
                    & GW GGA & 6.88 & 7.77 & 5.03 \\
\end{tabular}
\end{ruledtabular}
\end{table}

\subsection{Effective masses}
\label{mass}
\begin{table}
\centering
\caption{Effective mass  ( in the unit of free electron mass $m_{e}$)\label{tabmass}}
 \begin{ruledtabular}
 \begin{tabular}{lcc}
 & BeSiN$_2$ & BeGeN$_2$ \\ \hline
 $m^c_x$ &0.54 &0.44     \\
$m^c_y$ & 0.68      & 0.54   \\
$m^c_z$ &  0.36     & 0.33   \\ \\

$m^{a_1}_x$  &  2.59         &  2.00  \\
$m^{a_1}_y$  &   2.91        &   2.03 \\
$m^{a_1}_z$  &   0.25        &  0.22 \\
\\
$m^{b_1}_x$  &   0.37        & 0.26  \\
$m^{b_1}_y$  &  4.52         &  2.09  \\
$m^{b_1}_z$  &  2.33         &  1.89  \\
\\

$m^{b_2}_x$  & 5.77          &  2.15  \\
$m^{b_2}_y$  &  0.33         & 0.27   \\
$m^{b_2}_z$  &  2.54         &   1.88 \\

  \end{tabular}
  \end{ruledtabular}
\end{table}

\begin{table}
\centering
\caption{Parameters of effective Hamiltonian : inverse mass ($\hbar^2/m_e$)}\label{tabham}
 \begin{ruledtabular}
 \begin{tabular}{lcc}
      & BeSiN$_2$ & BeGeN$_2$ \\ \hline
$A_1$    & -4.00           &   -4.55                    \\
$A_2$    &       3.59           &  4.01                     \\
$A_3$    &     0.02             &  -0.001                     \\  
\\
$B_1$    &  -0.36                & -0.50                      \\
$B_2$    &    -1.17              &   -1.63                    \\
$B_3$    &    -0.07              &  0.04                     \\
\\
$C_1$    &    -0.02              &  -0.004                     \\
$C_2$    &    0.12              &    -0.03                   \\
$C_3$    &     1.33             &    1.65                   \\

  \end{tabular}
  \end{ruledtabular}
\end{table}

Finally, we determined the effective masses.  As mentioned in the computational
section, these include the $GW$ corrections to the bands. 
The CBM and VBM effective masses are given in Table \ref{tabmass}.
These correspond to the actual CBM between X and $\Gamma$ in BeSiN$_2$
and to the
CBM at $\Gamma$ in BeGeN$_2$. The effective masses here can be seen to be
significantly larger than in ZnGeN$_2$ or CdGeN$_2$ or even MgGeN$_2$.
As already mentioned, this corresponds to the lower dispersion of these
bands and the less deep Be-$s$ atomic energy levels on an absolute scale. 
The VBM is nearly degenerate but we give here the masses in the three
Cartesian directions for the three highest VB states.   We can see that for both
materials and for each state, there is one direction with a small mass and two with a large mass
$m_h>1.8$.  The direction of the small mass is the same in the two materials. 
In this case of orthorhombic crystal structure, the Cartesian axes coincide with crystal axes, i.e., $x=a, y=b, z=c$. The effective masses decrease from BeSiN$_2$ to BeGeN$_2$,  which is consistent with the trend found in other II-IV-N$_2$ compounds. The three upper valence bands can be best described by a Luttinger-like effective Hamiltonian.\cite{punya11} The inverse mass
parameters in such a Hamiltonian for the three top valence bands in BeSiN$_2$ and BeGeN$_2$ are derived from the masses given in Table \ref{tabmass} and
given in Table \ref{tabham}. 
\section{Conclusion}
\label{conclusion}
In this paper we considered the Be-IV-N$_2$ compounds, BeSiN$_2$ and
BeGeN$_2$, of which only the former has been synthesized in the past.
This study should be viewed in the context of the 
until now underexplored broader family of II-IV-N$_2$ semiconductors. 
We determined their optimized lattice parameters
and internal structural parameters
and found them to be in good agreement with previous DFT calculations
and the limited available experimental data. 
Their band structures were determined with the accurate and predictive
QS$GW$ method and indicate these materials have gaps in the deep UV. 
The indirect nature of the band gap of BeSiN$_2$ is confirmed at the QS$GW$ level. Effective masses were determined as well as band gap deformation potentials
and details of the valence band splittings, which are necessary
for future exploitation of these materials in heterostructure devices. 
In a band gap versus lattice constant plot, they fall in a quite well separated
area from the other III-N or II-IV-N$_2$ semiconductors, with a lattice
volume closer to those of the ultra wide band gap and extremely hard materials,
diamond and tetrahedrally bonded cubic and wurtzite BN.
We discussed the relation of
their band structure to that of wurtzite BN in  terms of band folding,
which helps to explain the indirect nature of the gap of
BeSiN$_2$ ad the location of its CBM at 2/3 $\Gamma-X$.

\acknowledgements{This work was supported by the National Science Foundation,
  Division  of Materials Research
under grant No. 1533957 and the DMREF program. 
Calculations made use of the High Performance Computing Resource in the Core Facility for Advanced Research Computing at Case Western Reserve University.}
\bibliography{Bib/dft,Bib/abinit,Bib/zgn,Bib/zsn,Bib/msn,Bib/cd4n,Bib/pockel,Bib/elastic,Bib/be4n,Bib/gw,Bib/lmto}

\end{document}